%Paper: hep-ph/9312230
%From: Paul Frampton <frampton@physics.unc.edu>
%Date: Mon, 6 Dec 1993 11:08:31 -0500 (EST)

\font\titlefont = cmr10 scaled \magstep2
\magnification=\magstep1
\vsize=20truecm
\voffset=1.75truecm
\hsize=14truecm
\hoffset=1.75truecm
\baselineskip=20pt

\settabs 18 \columns

\def\b{\bigskip}
\def\bb{\bigskip\bigskip}

\def\ce{\centerline}

\def\no{\noindent}

%$$\eqalign{
% put in lines of equations here, each ending in \cr
%}$$

%$$\eqalign{
%put in equations here ending each line with \cr
%} \eqno (1)$$
%the above will put the one between the two lines of equations and set it
%off to the right

% END BEGINNING FORMATS % BEGIN HEADER
\rightline{ UMDHEP 94-72}
\rightline{ IFP-487-UNC}
\rightline{ December 1993}
\ce{\titlefont{Partial Derivation}}
\ce{\titlefont{ of}}
\ce{\titlefont{ Transformation Properties of Quarks
and Leptons}}

\bb
\ce{\bf{P.H. Frampton\footnote{*}{\rm{ Work
supported by a  U.S. Department of Energy  Grant No. DE-FG05-85ER-40219}},~~}}

\ce{\it{ Institute of Field Physics,Department of Physics and Astronomy,}}
\ce{\it{  University of North Carolina, Chapel Hill, NC 27599}}

\ce{\bf{ R.N. Mohapatra\footnote\dag{\rm{ Work supported by
  the National Science Foundation Grant No. PHY-91-19745}},~~   }}

\ce{\it{ Department of Physics, University of Maryland,
College Park, MD 20742}}

\b
\ce{\bf Abstract}

\no
 Under the assumptions that $SU(3)_c\times U(1)_Y \times G^{\prime}$
with $G^{\prime}$ simple is a local symmetry group at high energies,
that color is parity-conserving, and the Y-charges are irreducible,
we show that anomaly constraints imply the minimal set of fermions is
fifteen in number.
Given this minimal set, we further show that $G^{\prime}$ must be $SU(2)$
and the unbroken gauge symmetry is {\it either} color {\it or} the product
of color with electric charge.

 \filbreak

In discussing the standard model[1], one usually postulates the
known set of fermions $(u^{\alpha}, d^{\alpha}, e, \nu )$ for each generation
( where the superscript $\alpha = 1, 2, 3$ denotes the color index ), then one
assumes $SU(3)_c \times SU(2)_L\times U(1)_Y$ to be the gauge group and
assigns the gauge quantum numbers for the fermions to match their
observed properties .
In view of the extraordinary success of this model, it is interesting to
ask whether it is possible to reproduce it starting from a more economical
set of assumptions.  Such results will not only shed light on the
fundamental structure of the standard model but may also be useful
in pointing the way towards new physics. The present Letter
is an attempt in that direction.

Our basic tool will be the freedom from Adler-Bell-Jackiw anomalies[2]
required for the consistent renormalization and unitarity of a gauge
field theory. Such considerations
have already been successfully used in the past[3] to show that there
is no need to pre-assign the Y-charges of the fermions of the
one-generation standard model but rather they can be determined by the
constraints of anomaly freedom once the fermion content
and the gauge group are given, if QED is assumed to be vector-like.
 Our goal here is to go further
using the same anomaly constraints. Our starting
assumptions are that the local symmetry of electroweak and strong
interactions at high energies is given by $SU(3)_c\times U(1)_Y\times
G^{\prime}$, that $SU(3)_c$ is vector-like and that the set of Y-charges
of the fermions is irreducible. The first two
assumptions are self-explanatory but the third needs to be explained.
By {\it irreducible} set of Y-charges we mean that no fermion has $Y=0$,
no pair of fermions have equal and opposite Y-charges, and more generally
{\it no subset of the fermions separately satisfies all of the anomaly
constraints}. One motivation
behind this assumption is that if a pair of fermions have equal and
opposite Y-charges and are vector-like or neutral under color and $G^{\prime}$,
 then one can form a gauge invariant mass term for
them and {\it a priori} this mass has no reason to be below the electroweak
scale implying that these fermions will decouple from the low energy
spectrum of the theory.

We find that the above assumptions
combined with the requirement that the gauge group be anomaly-free
leads to the following conclusions:

(i) The minimal number of fermions that leads to an
anomaly-free theory is 15, which is precisely the
number of fermions in the one-generation standard model;

(ii) The maximal allowed simple $G^{\prime}$ is $SU(2)$ which must be
parity-violating;

(iii) The unbroken gauge symmetry has to be {\it either} color alone {\it or}
the product of color with electric charge.

We believe that the set of assumptions we have made are more economical
than those made in the usual construction of the standard model and we are
able to reproduce three key ingredients of  the one generation
standard model i.e. the number of fermions, the weak gauge group and
the correct quantum numbers of the fermions.

In deriving these results we make use of the fact that charged chiral
fermions can acquire mass only if vector-like with respect to the corresponding
unbroken gauge symmetries. Gauge symmetries which are parity-violating
must generally be broken to avoid exact masslessness of charged chiral
fermions. This is why the electroweak gauge symmetry $SU(2)\times U(1)$
must be spontaneously broken {\it either} completely {\it or} to leave only
the $U(1)$ of electric charge.

In order to prove our assertion, let us say that the number of fermions is N
and is divided into two groups called quarks and leptons, the quarks
being defined as triplets or antitriplets under color and leptons being
singlets. The assumed vector nature of QCD requires the number of triplets and
antitriplets to be the same; let this number be equal to $Q/2$ where Q is
hence an even number. Denoting the number of leptons by $L$, one has $N=3Q+L$.
Let the leptons and quarks have Y-charges $y_i$ $(1\leq i \leq L)$ and
$z_j$ $(1 \leq j \leq Q)$ respectively. The three anomaly constraints
arising from  $U(1)[Gravity]^2$, $U(1)[SU(3)_c]^2$ and $[U(1)_Y]^3$
lead to the following equations:
$$\Sigma_1^L ~y_i = 0;\eqno(1)$$

$$\Sigma^Q_1 ~z_j = 0;\eqno(2)$$

$$\Sigma^L_1 ~y^3_i + 3~ \Sigma^Q_1~ z^3_j = 0;\eqno(3)$$

As mentioned,we  assume all Y-values to be rational
 and we will normalize them
such that all $y_i$ and $z_j$ are non-vanishing positive or negative
integers. We will be interested in finding the smallest N for which the
$y_i$ and $z_j$ will satisfy Eqs.(1)-(3).  The assumption of
irreducibility implies that $L\geq3$.
Further since Q is even it can be $2, 4,
6$ etc.
  If $Q = 2$, by Eq. (2) it has to be a reducible set and is excluded by
our assumptions. The smallest value of Q is therefore 4 leading to
$N = 15$. This proves our first assertion above. Note that in order
to derive this result, nowhere have we used the anomaly constraints for
the group $G^{\prime}$.

Let us now show that the maximal group $G^{\prime}$ is $SU(2)$.
First we show that $G^{\prime}\neq SU(3)$.
Since this $SU(3)$ must be orthogonal to color $SU(3)_c$, the three
leptons must be a triplet under it in which case, Eq.(1) cannot be
satisfied.  This leaves as the only possible simple group $G^{\prime}$=$SU(2)$.

As a brief digression, suppose $G^{\prime}$ were not simple but merely
a $U(1)$. We shall now show the extra anomaly constraints associated
with it imply that it is vectorlike.
We can always take
linear combinations of the two $U(1)$ charges to define two new $U(1)$'s
and call their charges X and Y respectively. By means of an appropriate
choice we can make it vanish for one of the leptons. Let us denote
the X-charges of quarks to be $x_a$ where $a = 1~~to~~4$ and those of
leptons to be $x_5, ~x_6~,0$ . The mixed $U(1)[Gravity]$ and
$U(1)[SU(3)]^2$ anomalies then imply :

$$x_1 + x_2 + x_3 + x_4 = 0;\eqno(4a)$$

$$x_5 =- x_6;\eqno(4b)$$

$$x^3_1+x^3_2+x^3_3+x^3_4=0;\eqno(4c)$$

Again as before if we choose the X charges also to be rational,
then Eq.(4), combined with the Fermat's last theorem will imply that
$x_1=-x_2$ and $x_3=-x_4$ i.e. $U(1)_X$ is parity conserving. We
are led directly to identify $G^{\prime}$ with electric charge
and there is no non-abelian nature to the ornamentation of
color - {\it there are no weak interactions}!

Thus the only parity-violating choice, like the only
simple-group choice, is $G^{\prime} = ~SU(2)$, as we henceforth assume.

The next question then arises : how do the quarks and leptons transform
under this $SU(2)$ group ? Consistent with our assumptions,
there can be at most one doublet among the quarks otherwise Eq.(2) will
imply that the Y-charges for the quarks become reducible in conflict with
our assumptions. So far we are not using the anomalies associated
with the $SU(2)$ group.

Let us now require that
there must be an $SU(2)$ doublet among the leptons.
{}From this we  conclude that two of the $y_i$'s must be equal.

Now Eq. (3) implies that

$$(z_1+z_2)(z_2+z_3)(z_3+z_1) = -2y^3/3;\eqno(5)$$

\noindent
where $y_1=y_2=y=-y_3/2$. Eq.(5) dictates that $y$ is divisible by three
and we may take as the simplest possibility $y_1=y_2=-3$ and $y_3=6$.
Thus

$$(z_1+z_2)(z_2+z_3)(z_3+z_1) = +18.\eqno(6)$$

There are four independent ways to factor 18 into three integers:

\noindent
$(1)(2)(9)$; $(1)(3)(6)$; $(1)^2(18)$ and $(2)(3)^2$. It can be
shown that these factorizations correspond to the sets

\noindent
$(z_i)$ = $(-3,+4,+5,-6)$; $(-1,+2,+4,-5)$; $(-8,+9,+9,-10)$
and

\noindent
$(+1,+1,+2,-4)$ respectively.
Only the last of these allows an unbroken
gauge symmetry beyond color. This implies that, there is only one
doublet of $SU(2)$ in the quark sector.

Given that the quarks contain an $SU(2)$ doublet we can write the Y
charges as follows:
$\left(y_1,y_1,-2y_1;z_1,z_1,z_2, -(2z_1 + z_2) \right)$. This notation
makes it clear which
particles are in $SU(2)$ doublets.In this notation, the anomaly
constraints in Eqs. (1) and (2) are automatically satisfied and Eq.(3)
becomes:

$${{y^3_1}\over{3}} = -z_1 (z_1 + z_2)^2;\eqno(7)$$

We will now show that vector-likeness of the unbroken $U(1)$
will then determine not only the charge formula but also the
individual Y-charges of the fermions. The generator of the
final unbroken $U(1)$ can be written as:

$$Q_e = T_3 + \eta Y;\eqno(8)$$

The constraints of vector-like $Q_e$ are:

$$-{{1}\over{2}}+\eta y_1 = 2\eta y_1;\eqno(9a)$$

$$+{{1}\over{2}}+\eta z_1 = -\eta z_2;\eqno(9b)$$

$$-{{1}\over{2}} + z_1 = +\eta (2z_1 + z_2);\eqno(9c)$$

Equation (9b) and (9c) are actually equivalent; taken together
Eqs.(9) imply that $\eta = -1/{2 y_1}$ and $ y_1 = z_1 + z_2 $.
Putting these relations in eq.(4), we get $ y_1 = -3 z_1$ and
$ z_2 = 2 z_1$. With the normalization $z_1 = 1$ one confirms
that the set of 15 integers $(1)^6(2)^3(-3)^2(-4)^3(6)^1$ is a
solution of Eqs. (1)-(3) which is irreducible in the sense
defined above.

If we now rewrite Y as $Y^{\prime} = - \eta Y$,
and call $Y^{\prime} $ as the standard model Y, then the electric
charge formula becomes

$$ Q_e = T_3 + {{Y}\over{2}}.\eqno(10) $$

The  values of the new Y are precisely those of the standard model.
Note also that in deriving the above electric charge formula nowhere
have we used any assumption about the Higgs structure of the theory.
Also, we have derived that if there is an unbroken gauge group
beyond color then it must be a $U(1)$ generated by $Q_e$ of Eq. (10).

In conclusion, we have given an alternative {\it partial derivation } of the
one generation
standard model starting from a rather economical set of assumptions.
We may restate our conclusion in the following words. It is taken as
given that the color gauge group $SU(3)_c$ is unbroken and vector-like
with an equal number of triplets and antitriplets. This QCD is
then ornamented by an additional would-be-electroweak gauge symmetry
$U(1)_Y\times G^{\prime}$ with simple $G^{\prime}$. Assuming the Y
charges to form an {\it irreducible} set we show that the minimal model
must  have  only 15 fermions matching those of the standard model.
We then find that $G^{\prime}$ is a gauged $SU(2)$.
Finally requiring the unbroken gauge group to be vectorlike
implies that it be color alone or the product of color with electric
charge. In the latter case, the transformation properties of quarks and leptons
are uniquely derived.

\bb
\filbreak

\bb

\ce{\bf References}
\b
\item{[1]}
S. L. Glashow, {\it Nucl. Phys.} {\bf 22}, 579 (1961).
S. Weinberg, {\it Phys. Rev. Lett.} {\bf 19}, 1264 (1967);
A. Salam, in {\it Elementary Particle Theory } ( edited by N. Swartholm)
Almquist and Forlag, Stockholm, (1968);
\item{[2]}S. L. Adler, {\it Phys. Rev.} {\bf 177}, 2426 (1969);
J. S. Bell and R. Jackiw, {\it Nuov. Cim.} {\bf 51A}, 47 (1969);
W. Bardeen, {\it Phys. Rev.} {\bf 184}, 1848 (1969).
\item{[3]}R. E. Marshak and C. Q. Geng, {\it Phys. Rev.}
{\bf D39}, 693 (1989); A. Font, L. Ibanez and F. Quevedo, {\it Phys.
Lett.} {\bf 228B}, 79 (1989);
 N. G. Deshpande, Oregon Preprint OITS-107 (1979);
R. Foot, G. C. Joshi, H. Lew and R. R. Volkas, {\it Mod. Phys. Lett.}
{\bf A5}, 95 (1990);
K.S. Babu and R. N. Mohapatra, {\it Phys. Rev. Lett.} {\bf 63}, 938 (1989);
J.Minahan, P. Ramond and R. Warner, {\it Phys. Rev. } {\bf D 41}, 715 (1990);
S. Rudaz, {\it Phys. Rev.} {\bf D41}, 2619 (1990).
E. Golowich and P. B. Pal, {\it Phys. Rev.}{\bf D 41}, 3537 (1990)
\bye